\documentclass[letterpaper,twocolumn,english,pra,showpacs]{revtex4-1}
\usepackage{graphicx}
\usepackage{bm}
\usepackage[linkcolor=blue]{hyperref}
\usepackage{amsmath,amssymb}
\usepackage{subfigure}
\usepackage{lipsum}
\usepackage{color}
\usepackage{hyperref}

\begin{document}

\title{Two Polarization Entangled Sources from the Same Semiconductor Chip}
\author{Dongpeng Kang}
\email{dongpeng.kang@mail.utoronto.ca}
\affiliation{The Edward S. Rogers Department of Electrical and Computer Engineering, Centre for Quantum Information and Quantum Control,
University of Toronto, 10 King\textquoteright{}s College Road, Toronto,
Ontario M5S 3G4, Canada.}

\author{Minseok Kim}
\affiliation{The Edward S. Rogers Department of Electrical and Computer Engineering, Centre for Quantum Information and Quantum Control,
University of Toronto, 10 King\textquoteright{}s College Road, Toronto,
Ontario M5S 3G4, Canada.}

\author{Haoyu He}
\affiliation{The Edward S. Rogers Department of Electrical and Computer Engineering, Centre for Quantum Information and Quantum Control,
University of Toronto, 10 King\textquoteright{}s College Road, Toronto,
Ontario M5S 3G4, Canada.}

\author{Amr S. Helmy}
\affiliation{The Edward S. Rogers Department of Electrical and Computer Engineering, Centre for Quantum Information and Quantum Control,
University of Toronto, 10 King\textquoteright{}s College Road, Toronto,
Ontario M5S 3G4, Canada.}

\date{\today }

\begin{abstract}
Generating nonclassical states of photons such as polarization entangled states on a monolithic chip is a crucial step towards practical applications of optical quantum information processing such as quantum computing and quantum key distribution. Here we demonstrate two polarization entangled photon sources in a single monolithic semiconductor waveguide. The first source is achieved through the concurrent utilization of two spontaneous parametric down-conversion (SPDC) processes, namely type-0 and type-I SPDC processes. The chip can also generate polarization entangled photons via the type-II SPDC process, enabling the generation of both co-polarized and cross-polarized entangled photons in the same device. In both cases, polarization entanglement is generated directly on the chip without the use of any off-chip compensation, interferometry or bandpass filtering. This enables direct, chip-based generation of both Bell states $(|H,H\rangle+|V,V\rangle)/\sqrt{2}$ and $(|H,V\rangle+|V,H\rangle)/\sqrt{2}$ simultaneously utilizing the same pump source. In addition, based on compound semiconductors, this chip can be directly integrated with it own pump laser. This technique ushers an era of self-contained, integrated, electrically pumped, room-temperature polarization entangled photon sources.
\end{abstract}

\maketitle

\section{Introduction}

\par Entangled photons are essential building blocks for optical quantum information processing, such as quantum computing (QC)~\cite{Ladd_Nature_2010} and quantum key distribution (QKD) \cite{Gisin_RMP_2002}. Conventionally, entangled photons have been generated using a myriad of techniques, most notably by using the process of spontaneous parametric down-conversion (SPDC) utilizing second order nonlinearities in crystals \cite{Christ_2013}. Properties such as brightness, scalability, compact form-factor and room temperature operation play key roles in enabling us to fully profit from entangled photon sources in applications such as QC and QKD. As such, the physics and technology of generating and manipulating entangled photons in monolithic settings have recently been topics of immense interest. Harnessing such effects in a monolithic form-factor also enables further incorporation of other photonic components that may be necessary for the aforementioned applications~\cite{O'Brien_NP_2009,Gaggero_APL_2010,Silverstone_NP_2014,Jin_PRL_2014}. This provided the drive that motivated the early work on implementing entangled sources in waveguides of crystals with appreciable second order nonlinearities such as Lithium Niobate~\cite{Suhara_LPR_2009}.
\par Realizing entangled photon sources in monolithic settings enables much more than the inclusion of numerous necessary components simultaneously: It can enable the direct generation of novel and useful photonic quantum states with specified properties, without moving parts, while benefiting from the accurate alignment of nano-lithography, precision of epitaxial growth and thin film deposition techniques. For example, monolithic platforms offer opportunities to provide photons that are entangled in one or several degrees of freedom simultaneously without the need for any extra component on the chip~\cite{Zhukovsky_OL_2011,Kang_PRA_2014}.  
In addition, monolithic sources can offer significant control over the spectral-temporal properties of the entangled photons with relative ease and high precision \cite{Abolghasem_OL_2010}. This in turn provides a powerful tool for tailoring the temporal correlation or the spectral bandwidth of the photon states. Such states can be of extremely short correlation times, which can enhance the accuracy of protocols for quantum positioning and timing~\cite{valencia1} and the sensitivity offered by quantum illumination~\cite{lloyd1}. The same integrated sources can generate states with extremely large temporal correlation times. This in turn leads to narrow spectral bandwidth, which can provide a more efficient atom-photon interface and improved sources for long-haul QKD~\cite{Narrowband_Sauge}.

\par The vast majority of the aforementioned applications use polarization entangled photon sources. Entanglement in the polarization degree of freedom has been the most widely utilized to implement entangled sources for experiments and applications that probe or exploit quantum effects. Photon pairs in polarization entangled sources need to be indistinguishable in every degree of freedom, except for polarization, which is challenging to achieve for states produced directly in waveguides \cite{Suhara_LPR_2009,Kaiser_NJP_2012,Arahira_OE_2013}. For photon pairs generated in a type-II process, in which the down-converted photons are cross-polarized, the birefringence in the group velocities of the modes, where the photons propagate, will cause a temporal walk-off between the pair, allowing polarization to be inferred from the photon arrival time. On the other hand, for photon pairs generated in a type-0 or type-I process, where the photons in a pair are co-polarized, there is a lack of two orthogonal polarizations necessary for polarization entanglement. As a result, most waveguide sources of photon pairs require an off-chip compensation setup \cite{Kaiser_NJP_2012} or an interferometer \cite{Arahira_OE_2013} to generate polarization entanglement, which increases the source complexity and decreases the system stability significantly.

\par Recently, several techniques have been demonstrated to generate polarization entangled photons from a monolithic chip \cite{Matsuda_SR_2012,Olislager_OL_2013,Orieux_PRL_2013,Horn_SR_2013}. The approaches which use spontaneous four-wave mixing (SFWM) in Si-based chips utilize integrated photonic components such as on-chip polarization rotators \cite{Matsuda_SR_2012} or 2D grating couplers \cite{Olislager_OL_2013}, and benefit from mature fabrication technologies. However, the indirect bandgap of Si presents significant challenges for further integration with the pump lasers. To this end, III-V semiconductor material systems offer an optimal solution in terms of functionality to tailor the dispersion and birefringence as well as monolithic integration with the pump lasers \cite{Bijlani_APL_2013,Boitier_PRL_2014}. Techniques using the counterpropagating phase-matching (PM) scheme \cite{Orieux_PRL_2013} and modal PM in Bragg reflection waveguides (BRWs) \cite{Horn_SR_2013} based on AlGaAs have been demonstrated. In the former case, however, the requirement of two pump beams with strictly controlled incidence angles and beam shapes imposes significant challenge for further integration, while in the latter case, the spectral distinguishability and walk-off due to modal birefringence compromises the quality of entanglement.

\par
\begin{figure}[tbh!]
\centering
\includegraphics[width=0.95 \columnwidth]{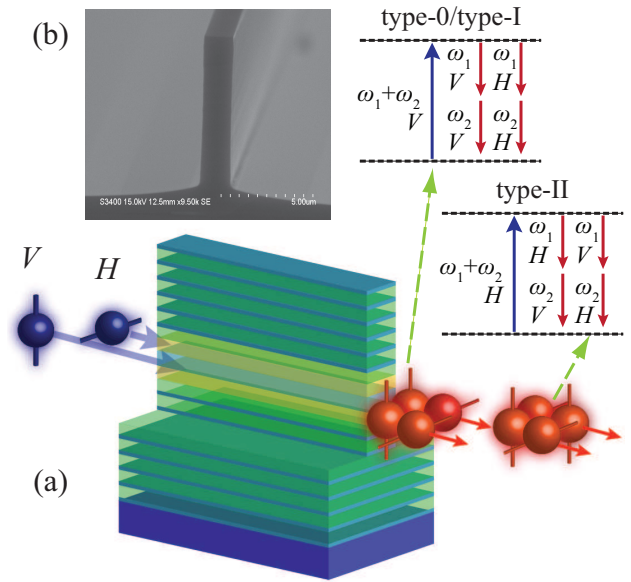}
\caption{(Color online) (a) The schematic of the multi-layer BRW as sources of polarization entangled photons. Entangled photons can be generated via concurrent type-0 and type-I processes, in which a TM polarized pump photon generates a pair of co-polarized photons, either both TE, or TM polarized. Instead, entangled photons can also be produced via the type-II process, in which a TE pump photon produces a pair of cross-polarized photons. Paired photons will be spatially separated, in this case, by a beamsplitter. The virtual energy diagrams of the two possible decaying channels for both entanglement generation mechanisms are also plotted. (b) An SEM image of the fabricated waveguide.}
\label{Fig:structure_SPDC_SEM}
\end{figure}

 In this work, we demonstrate how the waveguiding physics associated with BRWs can be used to simultaneously produce two polarization entangled photon sources using alternative approaches in a single self-contained, room-temperature semiconductor chip. The waveguide structure utilized is schematically shown in Fig. \ref{Fig:structure_SPDC_SEM}(a). The chip, based on a single monolithic semiconductor BRW, is straightforward to design and implement and has no moving parts. 
The technique allows direct polarization entanglement generation using an extremely simple setup without any off-chip walk-off compensation, interferometer, or even bandpass filtering. The first source is achieved through the concurrent utilization of two second order processes, namely type-0 and type-I SPDC processes, pumped by a single waveguide mode~\cite{Kang_OL_2012} as opposed to two modes of different polarizations~\cite{Matsuda_SR_2012} or modes propagating in opposite directions \cite{Orieux_PRL_2013}. 
This approach permits the integration of the pump with the source in a monolithic form. 
Within the same waveguide, there exists a second source based on type-II process due to the lack of material birefringence \cite{Horn_SR_2013}. The virtual energy diagrams of the two sources are also schematically shown in Fig. \ref{Fig:structure_SPDC_SEM}(a). As such, in this approach, by varying the pump polarization and wavelength, one can select between both polarization entangled sources or use them concomitantly. The direct generation of both Bell states $(|H,H\rangle+|V,V\rangle)/\sqrt{2}$ and $(|H,V\rangle+|V,H\rangle)/\sqrt{2}$ on a single chip can be envisaged. In addition, by lithographically tuning the waveguide ridge width, one can tune the degree of entanglement of the first source. 
\section{Methods}
\par For concurrent type-0 and type-I processes with a shared TM polarized pump, paired photons can be either generated in TM polarizations via the type-0 process, or in TE polarizations via type-I process. In the ideal case, photon pairs can be produced from the two processes with the same efficiency and identical spectrum, which renders them in a maximally entangled state $(|H,H\rangle+\exp{i\phi}|V,V\rangle)/\sqrt{2}$. This is the approach which we shall pursue to obtain a chip-based entangled source using the type-0 and type-I interactions. 
\par The AlGaAs structure used to demonstrate these sources was grown on a [001] GaAs substrate and the waveguide direction was oriented along [110]. Due to the zinc-blende crystal symmetry, the nonlinear tensor $\chi^{(2)}_{ijk}$ is non-zero only when $i\neq j\neq k$, with $i,j,k=x,y,x$ being the crystal coordinates. As a result, three SPDC processes, namely type-0, type-I, and type-II could coexist provided PM is satisfied. Among them, type-0 process depends on the electric field components of the interacting modes along the propagation direction, which are usually negligible in weakly guided waveguides. In BRWs, however, due to the index variations between different layers, the efficiency of type-0 process can be significant and can even be markedly tuned by engineering the epitaxial structure \cite{Kang_OL_2012}. In order to achieve concurrent PM of type-0 and type-I processes, the effective indices of the pump $n_{\text{TM}}(2\omega)$ should be equal to those of the down-converted photons $n_{\text{TE}}(\omega)$ and $n_{\text{TM}}(\omega)$ simultaneously, i.e., $n_{\text{TM}}(2\omega)=n_{\text{TE}}(\omega)=n_{\text{TM}}(\omega)$, with $\omega$ indicating the degenerate PM frequency of the down-converted photons, and TE, TM indicating the polarizations. This requirement can be satisfied lithographically by tuning the ridge width.

\par
\begin{figure}[tbh]
\centering
\includegraphics[width=0.99\columnwidth]{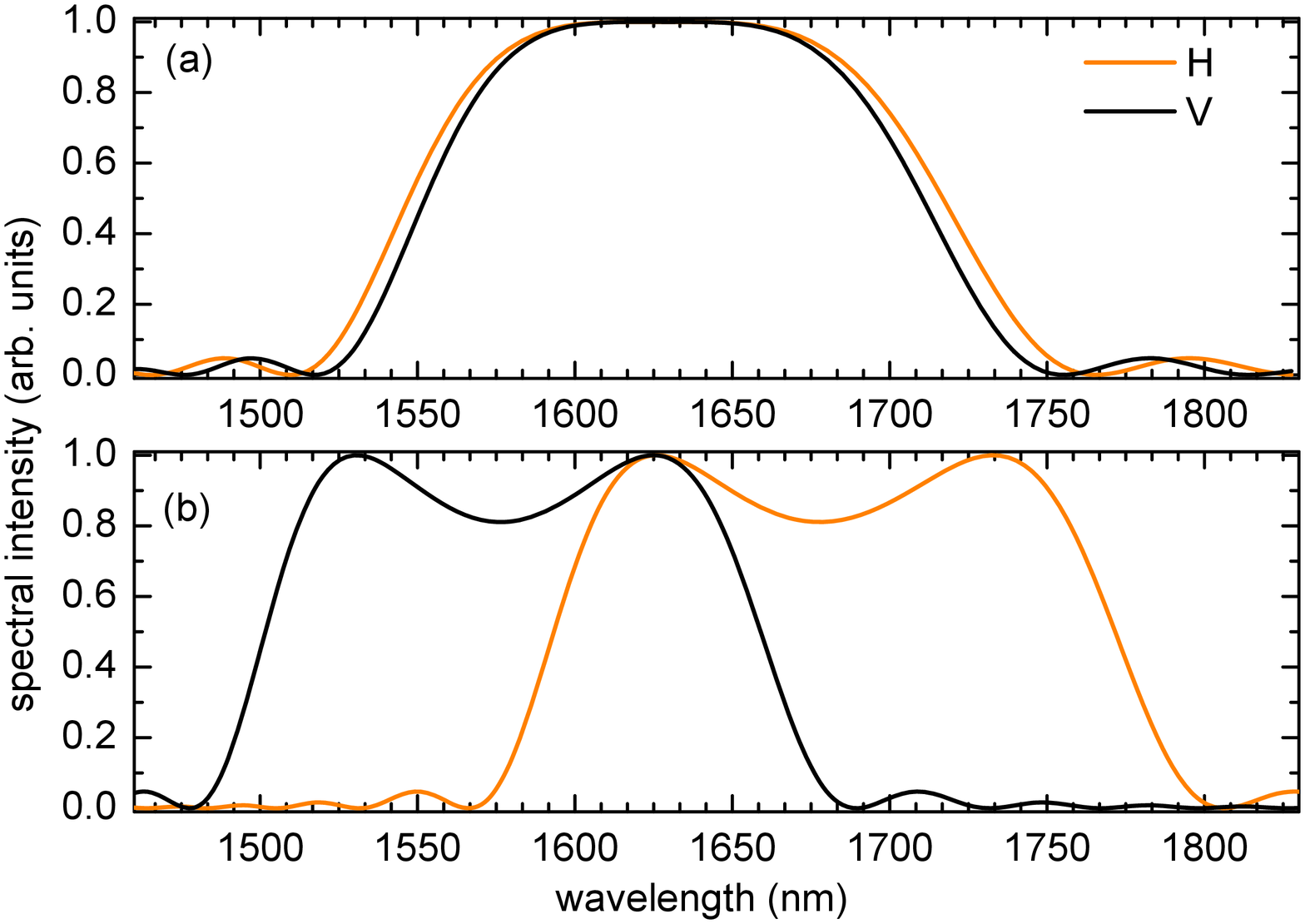}
\caption{(Color online) (a) The simulated spectral intensities of $H$ and $V$ polarized photons generated via the type-I and type-0 process, respectively; and (b) those of photons generated via the type-II process. The waveguide length is assumed to be 1.05 mm, the same as the waveguide tested in the experiment.}
\label{Fig:spectra}
\end{figure}

 The two photon state generated via the two concurrent SPDC processes is given by \cite{Kang_PRA_2014,Zhukovsky_JOSAB_2012}
\begin{align}
\left|\psi\right\rangle=&\frac{1}{\sqrt{\eta_{\text{I}}+\eta_0}}\iint d\omega_1 d\omega_2[\sqrt{\eta_{\text{I}}}\Phi_{HH}(\omega_1,\omega_2)|H\omega_1,H\omega_2\rangle \nonumber\\
&+\sqrt{\eta_0}\Phi_{VV}(\omega_1,\omega_2)|V\omega_1,V\omega_2\rangle],
\label{Eq:state}
\end{align}
where $\eta_{\text{I}}$, $\eta_{0}$ are the generation rates (pairs per pump photon) of the two processes after taking into account the losses. $\Phi_{HH}(\omega_1,\omega_2)$ and $\Phi_{VV}(\omega_1,\omega_2)$ are the biphoton wavefunctions, with the subscripts indicating the photon polarizations, and satisfy the normalization condition $\iint{d\omega_1 d\omega_2|\Phi_{HH(VV)}(\omega_1,\omega_2)|^2}=1$. When spatially separated, the paired photons are polarization entangled. The two spectra are found to be almost identical, as shown in Fig. \ref{Fig:spectra}(a). As a result, Eq. (\ref{Eq:state}) is maximally entangled when the generation rates are the same, i.e., $\eta_{\text{I}}=\eta_0$. In this case, there is no way, even in principle, to tell in which process a pair of photons are generated unless polarizations are measured. Therefore, polarization entangled photons can be generated inherently on the chip without the need of any extra component.

\par Polarization entanglement can also be produced by the type-II process on the same chip. Following the same formalism, the two-photon state of the type-II process can be explicitly written as
\begin{align}
\left\vert\psi^{\prime}\right\rangle=&\frac{1}{\sqrt{2}}\iint{d\omega_{1}d\omega_{2}}[\Phi_{HV}(\omega_1,\omega_2)|H\omega_1,V\omega_2\rangle \nonumber\\
&+\Phi_{VH}(\omega_1,\omega_2)|V\omega_1,H\omega_2\rangle].
\label{Eq:state_type-II}
\end{align}
For maximal entanglement, it requires $\Phi_{HV}(\omega_1,\omega_2)=\Phi_{VH}(\omega_1,\omega_2)$. This is not satisfied for the waveguide under test. However, due to the lack of material birefringence, and thus very small temporal walk-off, there exists a significant amount of overlap between $\Phi_{HV}(\omega_1,\omega_2)$ and $\Phi_{VH}(\omega_1,\omega_2)$. The corresponding spectra are shown in Fig. \ref{Fig:spectra}(b). As a result, entanglement exists even without any compensation and bandpass filtering.

\par Photons in a pair need to be spatially separated. In this work, we used a 50:50 beamsplitter to separate photons non-deterministically followed by post-selection. However, paired photons can also be separated deterministically by a dichroic mirror or integrated dichroic splitter as was done in \cite{Horn_SR_2013}. For an ideal dichroic mirror which has a splitting wavelength at the degenerate point, the degree of entanglement is identical to that using a 50:50 beamsplitter.

\section{Sample Description and Experimental Details}
\par As a proof of principle demonstration, a wafer designed for type-I PM around 1550 nm \cite{Abolghasem_OE_2010} is used to demonstrate this entangled photon source. Waveguides of this structure are lithographically tuned in order to satisfy concurrent PM of type-0 and type-I processes, with an etch depth of 6.5 $\mu$m and multiple ridge widths centered at 1.5 $\mu$m with a step size of 20 nm. An SEM image of a waveguide is shown in Fig. \ref{Fig:structure_SPDC_SEM}(b). Numerical simulations predict that concurrent PM of the two types can be achieved at a wavelength around 1.63 $\mu$m with a ridge width of $\sim$1.5 $\mu$m. Note that redesigning the epitaxial structure can shift the center wavelength to 1550 nm \cite{Kang_OL_2012}. The sample under test had a length of 1.05 mm. 
\par
\begin{figure}[tbh]
\centering
\includegraphics[width=0.99 \columnwidth]{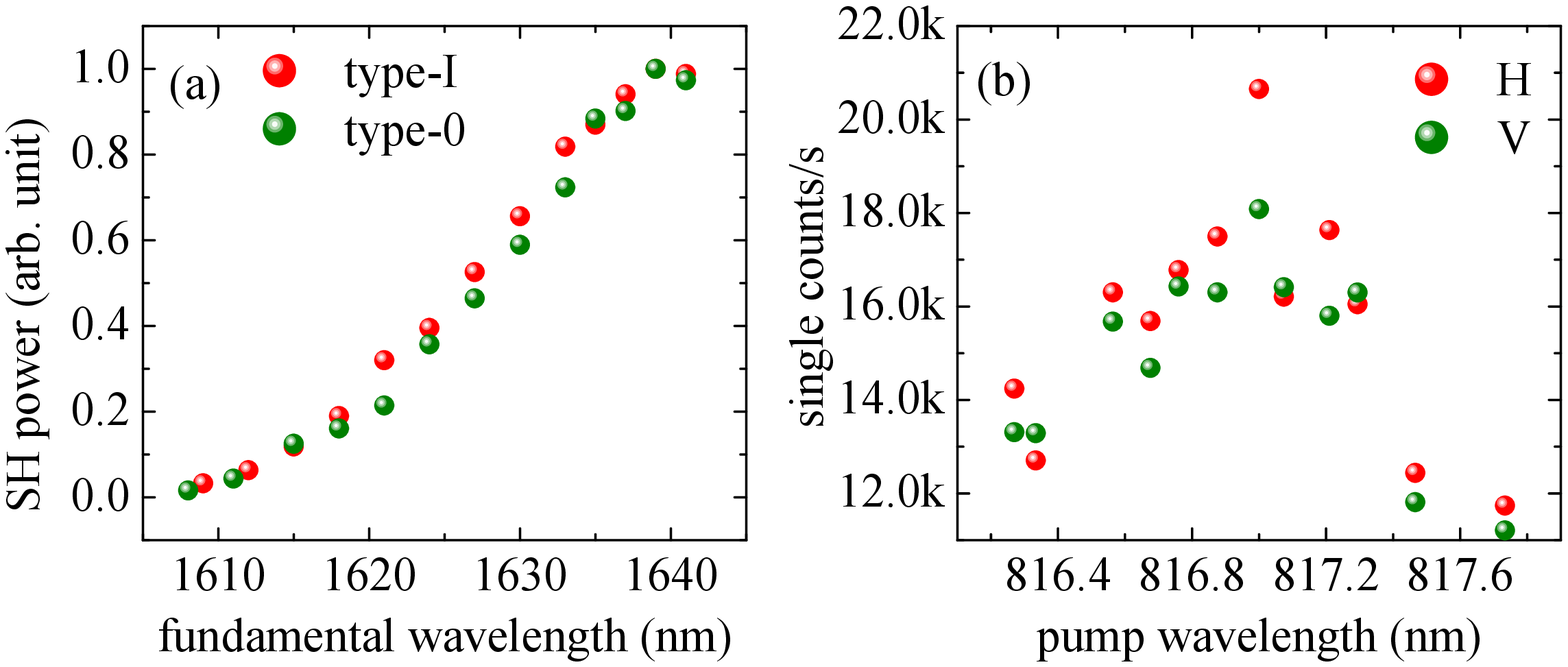}
\caption{(Color online) (a) The normalized SHG tuning curves for type-0 and type-I processes of the waveguide under test, and (b) the single counts versus pump wavelength for the $H$ and $V$ polarized photons generated from type-I and type-0 processes, respectively, with a fixed pump power. For the tuning curves shown in (a), the maximal wavelength is limited to the longest available wavelength of the pump OPO used in the experiment ($\sim$1640 nm).}
\label{Fig:PM wavelength}
\end{figure}

 In order to select the waveguide that has the best alignment of their PM wavelengths, second harmonic generation (SHG) for both type-0 and type-I processes were tested. The experiment was carried out on a standard end-fire coupling setup by pumping the waveguides with an optical parametric oscillator (OPO) pumped by a femtosecond pulsed Ti:Sapphire laser. The normalized SHG tuning curves of the waveguide under test are shown in Fig. \ref{Fig:PM wavelength}(a). According to Fig. \ref{Fig:PM wavelength}(a), PM wavelengths of both types are near 1640 nm, the longest achievable wavelength of the OPO used. Due to the large bandwidth of the pump pulse, the exact PM wavelengths could not be accurately identified. SPDC was then carried out by pumping the waveguides with a CW Ti:Sapphire laser where the dependence of single photon count rate on the pump wavelength was measured. This allowed us to locate the waveguide which has the best overlap in PM wavelengths among all tested waveguides. The results of the waveguide selected, shown in Fig. \ref{Fig:PM wavelength}(b), indicates that the PM wavelengths of both types are $816.7\pm0.3$ nm. The uncertainty in the PM wavelength measurement was due to the pump power fluctuation in the waveguide because of Fabry-–P\'erot resonances, which could not be resolved with instrumentation available.
\par
\begin{figure}[tbh]
\centering
\includegraphics[width=0.99 \columnwidth]{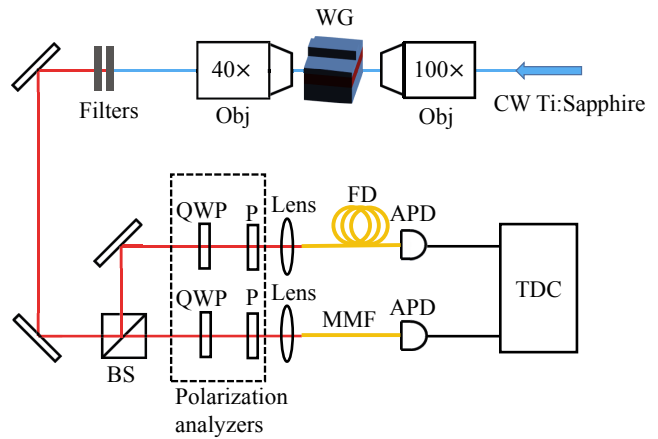}
\caption{(Color online) The schematic of the experimental setup for polarization entangled photon generation. The waveguide was pumped by a CW Ti:sapphire laser in an end fire setup, with a 100$\times$ input objective lens and a 40$\times$ output objective. Paired photons generated were separated by a 50:50 beamsplitter and passed though a pair of polarization analyzers consisting of a QWP and a polarizer. A pair of multimode fiber coupled InGaAs single photon detectors were used to detect the photons and the coincidence counting histograms were recorded by a TDC. Obj: objective; WG: waveguide; BS: beamsplitter; QWP: quarter-wave plate; P: polarizer; MMF: multimode fiber; FD: fiber delay; APD: InGaAs avalanche photo-diode; TDC: time-to-digital converter.}
\label{Fig:setup}
\end{figure}

 To generate polarization entangled photons via concurrent type-0 and type-I processes, the TM polarized pump beam from a CW Ti:sapphire laser was coupled into the waveguide using a 100$\times$ (N.A.=0.90) objective lens, with a power of 1.13 mW before the lens. Photon pairs generated were collected by a 40$\times$ objective lens at the output facet and passed through long pass filters to eliminate the pump. After their separation using a 50:50 beamsplitter, the signal and idler photons were collected into multimode fibers and detected by two InGaAs single photon detectors. The signal arm detector (ID220, ID Quantique) operates in a free-running mode and provided 20\% efficiency at 1550 nm. The idler detector (ID210, ID Quantique) was operating in a gated mode and provided 25\% at 1550 nm. It was externally triggered by the detection events of the first detector. An optical delay was added before the second detector to compensate for the electronic delay between the two detectors. Both detectors had a deadtime of 20 $\mu$s. The coincidence counts were recorded with the help of a time-to-digital converter (TDC) circuitry. At the degenerate wavelength of $\sim$1635 nm, the detection efficiencies are around 4\% and 5\%, respectively. A pair of quarter-wave plates (QWPs) and polarizers were used to measure the polarizations of the photon pairs. The schematic of the experimental setup is illustrated in Fig. \ref{Fig:setup}.
\par Considering the transmission coefficients of the output objective lens (90\%), long pass filters (70\%), beamsplitter (43\%), QWPs and polarizers (75\%) and fiber collection efficiencies in each path (53\% and 34\%), the overall collection efficiency of photon pairs with respect to the position right after the waveguide output facet was found to be $\sim$1.5\%. 
\section{Results and Discussion}
\par
\begin{figure}[tbh]
\centering
\includegraphics[width=0.95\columnwidth]{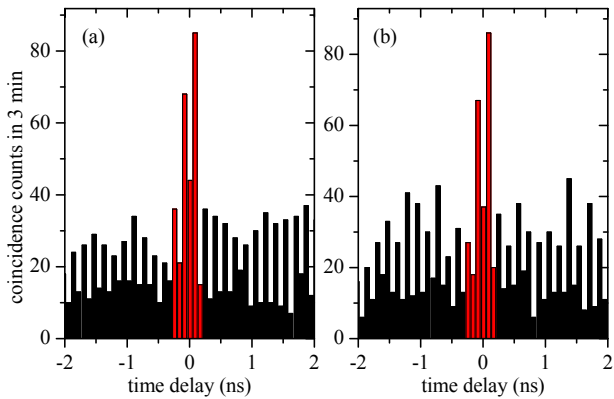}
\caption{(Color online) Coincidence histograms of photon pairs in the (a) $HH$ and (b) $VV$ basis for an integration time of 3 minutes. The red bars around the peaks represent the counts in the coincidence window of $\sim$0.5 ns.}
\label{Fig:histograph}
\end{figure}

 Typical coincidence histograms are given in Fig. \ref{Fig:histograph}(a) for two $H$ polarized photons and Fig. \ref{Fig:histograph}(b) for two $V$ polarized photons for a pump wavelength of 816.76 nm and an integration time of 3 minutes. The coincidence peaks indicate that photon pairs were generated via both type-I and type-0 processes. The high level of accidental counts is due to detector dark counts and broken photon pairs due to waveguide losses as well as limited collection (1.5\%) and detection (0.2\%) efficiencies. By blocking the idler arm, we found the dark counts of the second detector consists of 83\% of the total accidental counts. Thus we can expect a much higher coincidence-to-accidental (CAR) ratio by redesigning a sample which generates photon pairs in region where the detectors are more efficient (e.g. at 1550 nm). 
\par The net coincidence count rates for both type-0 and type-I processes are around 0.7 Hz, after subtracting the accidental counts. Taking into account the signal arm detector's dead time, single count rate (16 kHz), as well as the overall collection and detection efficiencies, we estimate the photon pair generation rates after the waveguide are $3.4\times10^4$ pairs/s. The input objective lens has a transmission of 70\% and the coupling efficiency into the pump Bragg mode is estimated to be 6\%, resulting in an internal pump power of 47.3 $\mu$W right after the input facet. Therefore, the photon pair production rates are both around $7.3\times10^5$ pairs/s/mW with respect to the internal pump power and external photon pairs, or equivalently, $1.8\times10^{-10}$ pairs/pump photon. The fact that both processes have roughly the same generation rates, as opposed to type-I being more efficient than type-0 according to theoretical calculation, could be because that the TM mode has a smaller loss than that of the TE mode in this type of deeply etched waveguides. We confirmed this by measuring the losses using Fabry-–P\'erot method and found the losses are 4.3 cm$^{-1}$ and 2.5 cm$^{-1}$ for TE and TM modes, respectively. It could also be because the pump wavelength is slightly detuned from the degenerate PM wavelength of the more efficient process, as the two processes may not have exactly identical PM wavelengths.

\par
\begin{figure}[th]
\centering
\subfigure{\includegraphics[width=0.49\columnwidth]{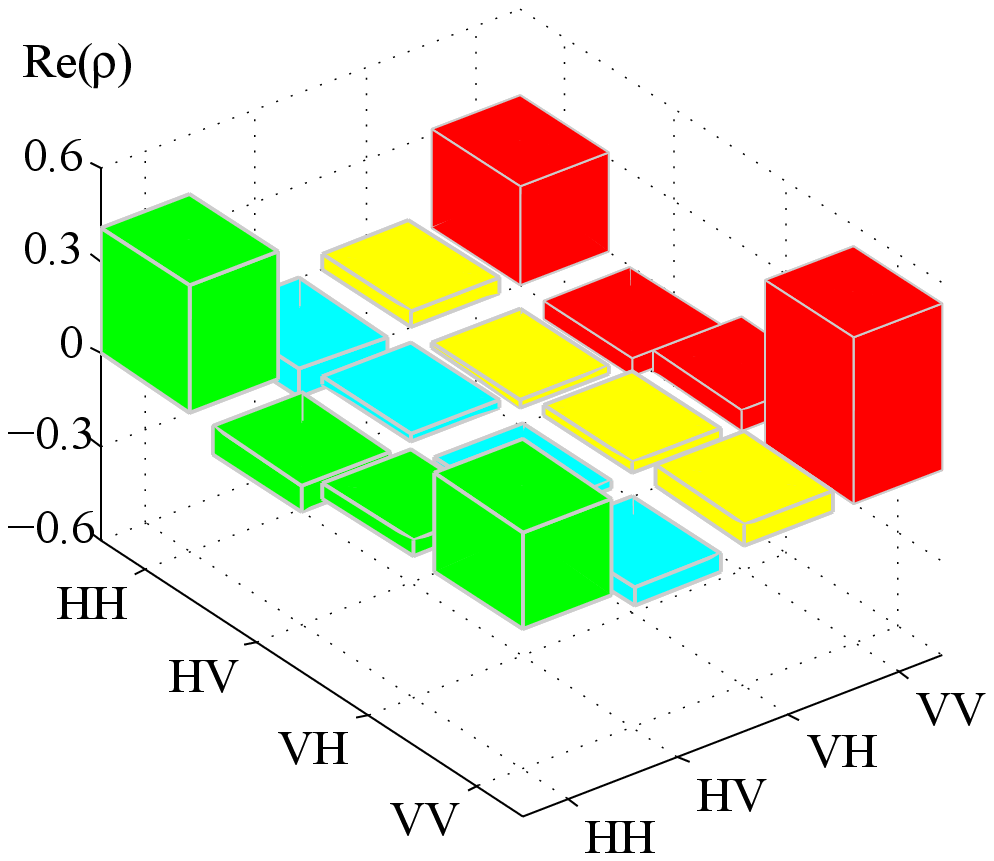}\label{Fig:rho_r}}
\subfigure{\includegraphics[width=0.49\columnwidth]{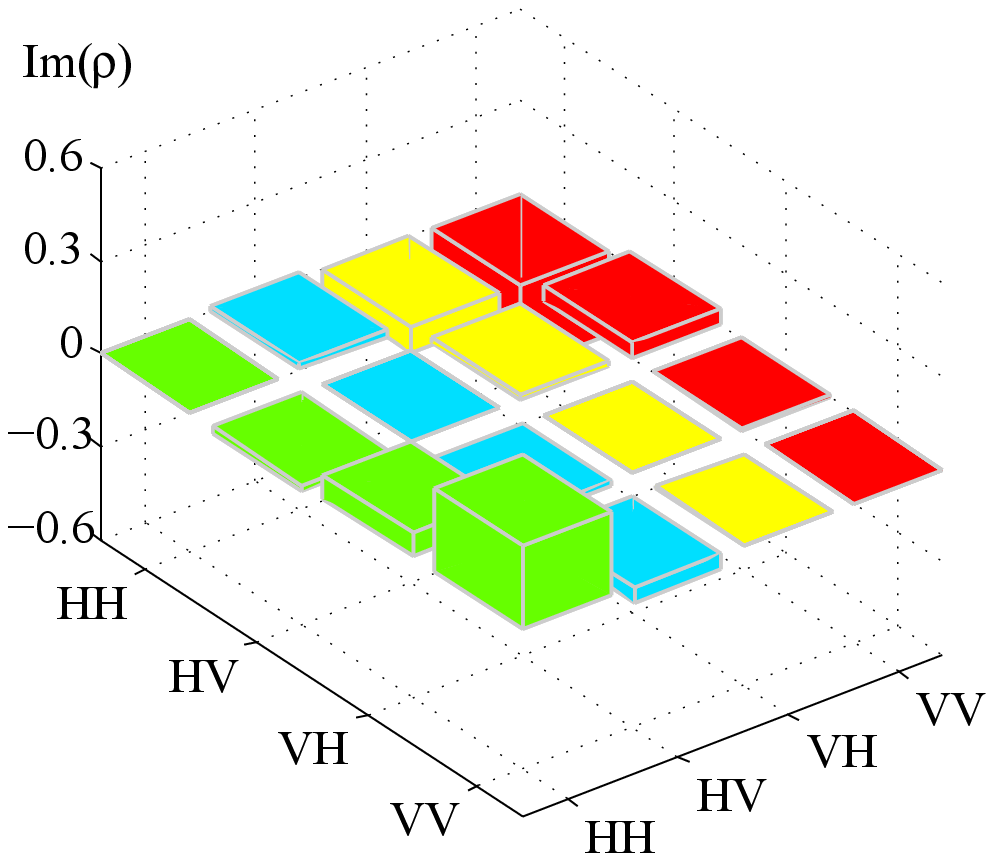}\label{Fig:rho_i}}
\caption{(Color online) The real and imaginary parts of the reconstructed density matrix of the output two-photon state via concurrent type-0 and type-I processes.}
\label{Fig:rho}
\end{figure}

 Quantum state tomography measurements were then subsequently performed by projecting the paired photons into 16 polarization combinations and the density matrix was reconstructed using the maximal likelihood method \cite{James_PRA_2001,note2}. The net density matrix $\rho$ of the output two-photon state, given by Fig. \ref{Fig:rho}, is found to have a concurrence \cite{Wootters_PRL_1998} of $0.85\pm 0.07$. The maximum fidelity with a maximally entangled state $|\Phi\rangle=(|H,H\rangle+\exp{(i\phi)}|V,V\rangle)/\sqrt{2}$, defined by $F=\max_{\phi}{\langle\Phi|\rho|\Phi\rangle}$, is 0.89 with a corresponding phase angle $\phi=40^{\circ}$. The non-zero phase $\phi$ is because of the slightly dissimilar degenerate PM wavelengths of the two processes. Theoretical calculation according to Eq. (\ref{Eq:state}) shows that this phase value is due to the type-I PM wavelength being $\sim$0.02 nm shorter than that of the type-0 process. The imperfection of the entanglement could be mainly because the pump wavelength is not optimal, causing extra spectral distinguishability between the two processes. This can be improved by using a tunable diode laser which has a fine spectral tunability. In addition, the mechanical drift of the characterization setup could result in an increase of mixture and a decrease of entanglement for measurements longer than a few minutes.
\begin{figure}[tb]
\centering
\includegraphics[width=0.95\columnwidth]{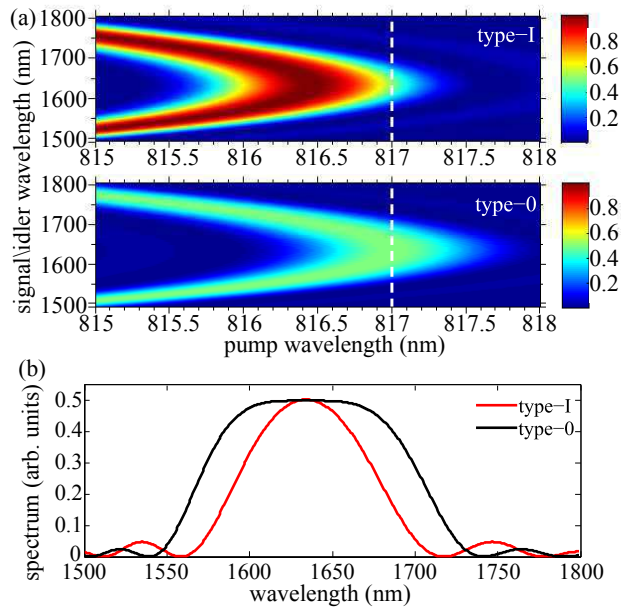}
\caption{(Color online) (a) The turning curves of the type-I (top) and type-0 (bottom) processes weighted by the corresponding efficiencies. The white dashed lines represent the pump wavelength, and (b) the corresponding spectra of the down-converted photons.}
\label{Fig:tuning_curves_and_spectra}
\end{figure}
\par The fact that a slightly different PM wavelengths causes one of the processes to take place below its maximal generation rate can be used to balance the generation rates of the two processes and therefore increase the degree of entanglement. Although engineering efforts can be made to achieve almost identical efficiencies, in reality, there may still be differences especially with the existence of polarization dependent losses. In such a case, the waveguide ridge width can be lithographically tuned such that the stronger process has a shorter PM wavelength, while the pump wavelength should be tuned to the degenerate PM wavelength of the weaker process. 

\par To illustrate this point, we consider an example where the type-I process is twice efficient as the type-0 process, i.e., $\eta_{\text{I}}=2\eta_{0}$. In the case where the two PM wavelengths are identical, the theoretical fidelity to a maximally entangled state is 0.96, and the concurrence is 0.93. The degree of entanglement can be increased to near maximum by tuning the ridge width such that the PM wavelength of type-I process is 0.55 nm below that of type-0 process. The corresponding tuning curves of the two processes weighted by their efficiencies are shown in Fig. \ref{Fig:tuning_curves_and_spectra}(a). The pump wavelength, marked by the white line, is fixed at the degenerate PM wavelength of type-0 process. In this case, the spectra, given by Fig. \ref{Fig:tuning_curves_and_spectra}(b), show almost the same amplitude, indicating the two processes having the same spectral brightness near the degenerate wavelength. By utilizing a weak spectral filtering with a bandwidth of 80 nm, the calculated fidelity and concurrence are increased to 0.997 and 0.995, respectively, with $\phi=34^{\circ}$, indicating the generation of maximally entangled photons. Without bandpass filtering, the maximum fidelity and concurrence can still reach 0.99 and 0.97, respectively, with a phase $\phi=14^{\circ}$, when the type-I PM wavelength is 0.36 nm below that of type-0 process.
\par In addition, this tuning approach could be used to generate non-maximally entangled states $(|H,H\rangle+r\exp{(i\phi)}|V,V\rangle)/\sqrt{1+r^2}$ with a tunable value of $r$, which offers significant advantages over maximally entangled states in some applications such as closing the detection loophole in quantum nonlocality tests \cite{Christensen_PRL_2013}.


\par
\begin{figure}[th]
\centering
\subfigure{\includegraphics[width=0.49\columnwidth]{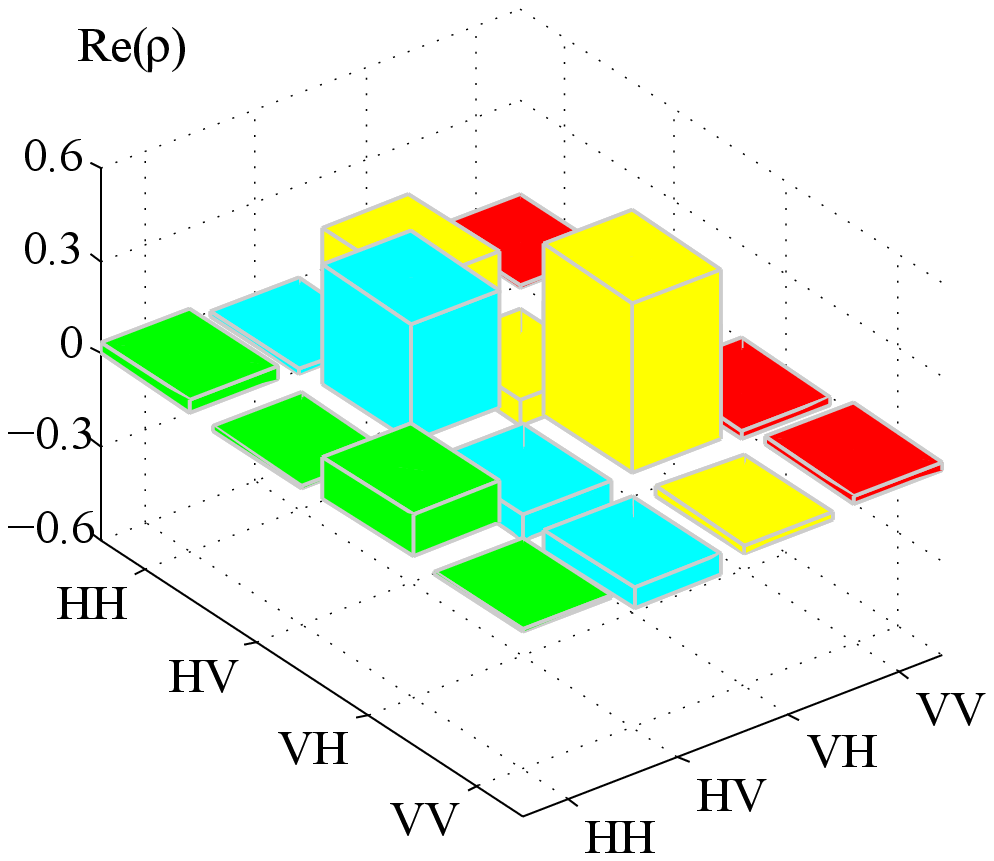}\label{Fig:rho_r_type-II}}
\subfigure{\includegraphics[width=0.49\columnwidth]{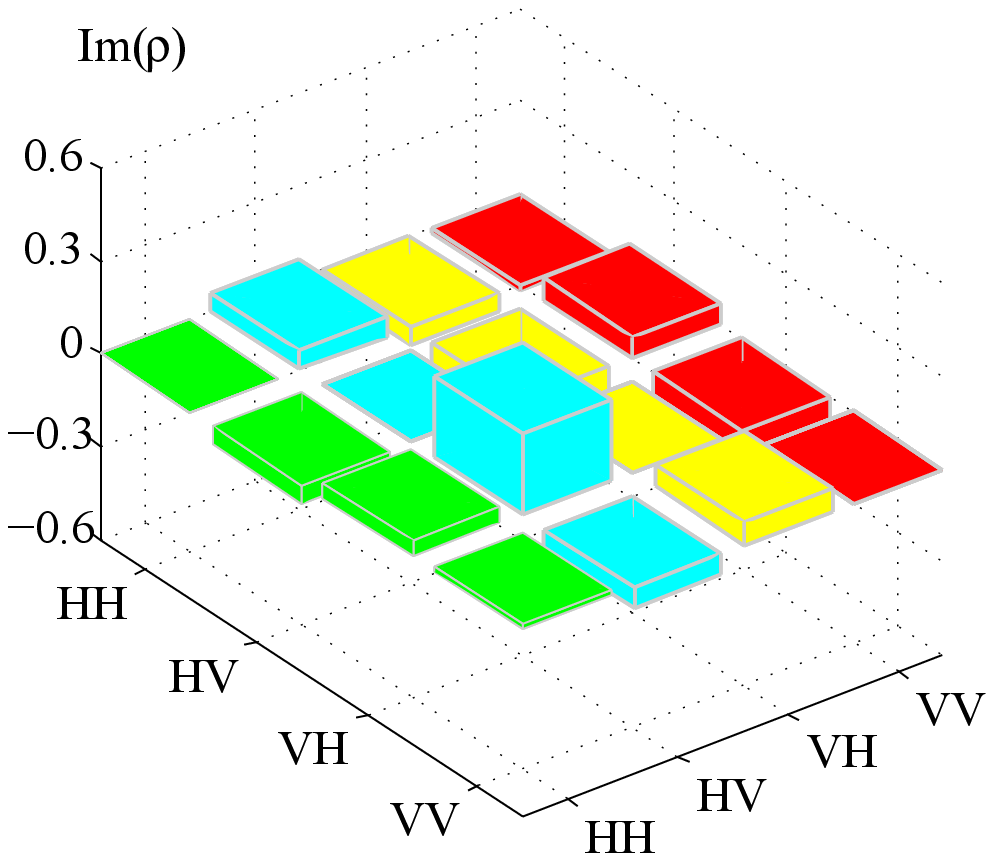}\label{Fig:rho_i_type-II}}
\caption{(Color online) The real (left) and imaginary (right) parts of the reconstructed density matrix of the output two-photon state generated via the type-II process.}
\label{Fig:rho_type-II}
\end{figure}

 Lastly, we show the generation of cross-polarized entangled photons on the same chip via the type-II process. For a TE polarized pump at 812.92 nm, which is only less than 4 nm below those of type-0 and type-I, the output state, again without any compensation and bandpass filtering, shows a concurrence of $0.55\pm 0.08$ and a fidelity of 0.74 to the maximally entangled state $(|H,V\rangle+\exp{i\phi}|V,H\rangle)/\sqrt{2}$. The degree of entanglement is comparable with that obtained in \cite{Horn_SR_2013}. The reconstructed density matrix is given by Fig. \ref{Fig:rho_type-II}. The utility of this device becomes most apparent when one considers its capability to have all three types of PM simultaneously achieved at the same wavelength via the tuning of the epitaxial structure combined with ridge width control \cite{Zhukovsky_OC_2013}. The unique ability to achieve this monolithically allows the generation of both co-polarized and cross-polarized polarization entangled photons on the same chip at the same pump wavelength, by simply selecting the pump polarization. Nevertheless, a few nanometres' difference in the PM wavelength can be covered by a femtosecond pump laser. 

\section{Conclusion}
\par In conclusion, we have demonstrated how the waveguiding physics associated with BRWs can be used to simultaneously produce two polarization entangled photon sources using alternative approaches in a single self-contained, room-temperature semiconductor chip. Direct generation of polarization entangled photons from a monolithic compound semiconductor chip via concurrent type-0 and type-I SPDC processes has been characterized. Simultaneous PM of the two processes was achieved using simple lithographic control on the ridge width of BRWs. Without the need of off-chip compensation, interferometer, and bandpass filter, the degree of entanglement is among the highest in previous demonstrations from monolithic III-V and Si chips. In addition, the same device can also directly generate polarization entanglement via the type-II process, with a pump wavelength of only 4 nm shorter.

\par Further improving the device performance relies largely on improved fabrications. By reducing the waveguide losses, the generation rates can potentially be increased by more than two orders of magnitudes, as predicted in \cite{Zhukovsky_JOSAB_2012} for lossless waveguides. In addition, the degree of entanglement can be increased by fine-tune the ridge width via more precise fabrications, as we have shown that the entanglement can be nearly maximal in the ideal case \cite{Kang_OL_2012}. The degree of entanglement can also be easily increased by bandpass filtering \cite{note}.

\par Note that previous work on ferro-electric waveguides also studied the diversity of multiple PM processes to generate quantum states of particular properties, such as using two quasi-phase matching (QPM) gratings to generate polarization entangled photons \cite{Herrmann_OE_2013,Gong_PRA_2011} and using different spatial modes to achieve mode entanglement \cite{Mosley_PRL_2009}. While they are fundamentally important, realizing such effects in monolithic fashions, especially on III-V semiconductor platforms as in this work, ushers a new era of practical optical quantum information processing.

%
\section*{Acknowledgements}
%
The authors would like to acknowledge R. Marchildon, G. Egorov, F. Xu, E. Zhu, and Z. Tang for helpful discussions. This work was supported by Natural Sciences and Engineering Research Council of Canada (NSERC).





\begin{thebibliography}{99}
\bibitem{note} In a separate experiment, we have demonstrated a concurrence of 0.98 via the type-II process in a similar waveguide with bandpass filtering only. The results will be published somewhere else.
\end{thebibliography}


\end{document}